\documentclass[3p,times,twocolumn]{elsarticle}

\usepackage{graphicx,amssymb,amsmath}
\usepackage{xspace}
\usepackage{url}
\usepackage{color} 
\usepackage{algorithm}
\usepackage{algorithmic}
\usepackage{tabularx}
\usepackage{subcaption}

\graphicspath{{img/}} 

\journal{Journal of Web Semantics}

\newdefinition{definition}{Definition}

\newcommand{\todo}[1]{} 
\newcommand{\quot}[1]{``#1''}
\newcommand{\sameas}{\texttt{owl:sameAs}\xspace}
\newcommand{\ld}{Linked Data\xspace}
\newcommand{\di}{data integration\xspace}
\newcommand{\odcs}{ODCleanStore\xspace} 
\newcommand{\odcsft}{ODCS\mbox{-}FusionTool\xspace}
\newcommand{\paygo}{pay-as-you-go\xspace}

\newcommand{\fq}{F\mbox{-}quality\xspace}
\newcommand{\FQ}{F\mbox{-}Quality\xspace}
\newcommand{\ngraph}{Named Graph\xspace}
\newcommand{\oi}{duplicate detection\xspace}

\newcommand{\aggreg}[1]{\textsc{#1}\xspace}
\newcommand{\power}[1]{\mathcal{P}(#1)}
\newcommand{\bigo}{{\mathcal{O}}}
\newcommand{\term}[1]{\textit{#1}}
\newcommand{\dbnull}{\texttt{NULL}\xspace}

\newcounter{listingcounter}
\newenvironment{customalgo}[2] {
  \begin{figure}[h!]
  \refstepcounter{listingcounter}
  \label{#1}
  \hrule
  \smallskip
  {\raggedright \textbf{Listing\ \thelistingcounter:} #2\par}
  \smallskip
  \hrule
  \smallskip
  \begin{algorithmic}[1]
} {
  \end{algorithmic}
  \smallskip
  \hrule
  \end{figure}
}

\newcounter{examplecounter}
\newcommand{\exampleref}[1]{Example~\ref{#1}}
\newenvironment{examplelisting}[2]{
  \begin{figure}[h!]
  \refstepcounter{examplecounter}
  \label{#1}
  \hrule
  \smallskip
  {\raggedright \textbf{Example\ \theexamplecounter}\ifx&#2&{}\else\textbf{:} #2\fi\par}
  \smallskip
  \hrule
  \smallskip
  \setlength{\parindent}{0pt}
  \setlength{\parskip}{0pt}
  \small
} {
  \smallskip
  \hrule
  \end{figure}
}

\newenvironment{customitemize} {
  \begin{enumerate}[\textbullet]
} {
  \end{enumerate}
}

\begin{document}

\begin{frontmatter}

\title{Linked Data Integration with Conflicts}

\author[]{Jan Michelfeit\corref{C}}
\ead{michelfeit@ksi.mff.cuni.cz}

\author{Tom\' a\v s Knap}
\ead{knap@ksi.mff.cuni.cz}

\author{Martin Ne\v cask\' y}
\ead{necasky@ksi.mff.cuni.cz}
\address{XML and Web Engineering Research Group, Charles University in Prague,
Malostransk\' e N\' am\v est\' i 25, 118 00 Praha 1, Czech Republic}

\cortext[C]{Corresponding author. Tel: +420 732 737 415}


\begin{abstract}
Linked Data have emerged as a successful publication format and one of its main strengths is its fitness for integration of data from multiple sources. This gives them a great potential both for semantic applications and the enterprise environment where data integration is crucial. 
Linked Data integration poses new challenges, however, and new algorithms and tools covering all steps of the integration process need to be developed.
This paper explores \ld integration and its specifics. We focus on data fusion and conflict resolution: two novel algorithms for Linked Data fusion with provenance tracking and quality assessment of fused data are proposed. The algorithms are implemented as part of the \odcs framework and evaluated on real Linked Open Data.
\end{abstract}


\end{frontmatter}

\section{Introduction}
\label{introduction}
More and more valuable datasets are being published on the Web and often their usefulness increases dramatically when data can be combined from multiple sources. Similarly, data in the enterprise environment are often distributed across many independent systems and their full value and potential can be exploited only when integrated together. Linked Data are an instrument specifically designed to facilitate linking and data integration across datasets and provide many advantages for data integration over alternative approaches. 

Data integration in relational databases is a well explored field with mature tools and frameworks covering all steps of integration. On the other hand, Linked Data integration has still open challenges.
One is the resolution of conflicts and uncertainties emerging during integration.
 Another challenge is information quality whose importance is significant especially in the open Web environment. Both data consumer and conflict resolution tools need support in decisions about which data are worth using.
  We must also face technical challenges -- integration must be efficient and a \paygo approach~\cite{paton2012pay} may be essential for adoption in practice.

\medskip 
This paper aims to fill in the missing pieces in the data fusion step of Linked Data integration, and  address the outlined challenges. We provide both a practical tool and a theoretical framework as a basis for further research. The main contributions are:

\begin{enumerate}
  \item An algorithm realizing the data fusion step in \ld integration with resolution of conflicts and provenance tracking. The algorithm deals with usage of different identifiers and schemata, and conflicting or missing values in data sources.
  \item A conflict-based quality assessment algorithm which leverages information available during data fusion. We introduce the concept of \term{\fq} as a measure of quality of fused data as opposed to quality of source data.
  \item Overview of the specifics of \ld integration in comparison with relational databases.
\end{enumerate}

We show how the proposed algorithms work together to improve both the abilities of conflict resolution and quality assessment. Both algorithms are implemented as part of a Linked Data integration framework \odcs and evaluated on real Linked Open Data.

The paper is structured as follows: Section~\ref{sec:di} examines the data integration and data fusion process for Linked Data and introduces the \odcs framework. Section~\ref{sec:relatedWork} gives an overview of related work. Section~\ref{sec:cr} describes in detail the proposed data fusion algorithm and Section~\ref{sec:qa} covers conflict-based quality assessment. Section~\ref{sec:evaluation} presents experimental results. Section~\ref{sec:specifics} compares the specifics of \ld integration to relational databases and we summarize our results in Section~\ref{sec:conclusion}.


\section{Data Integration \& \odcs Framework}

\label{sec:di}
Data integration is about combining data from different sources to a unified view. The main challenges that must be addressed in order to achieve a unified integrated view include:

\begin{customitemize} 
  \item Technical and semantical heterogeneity of data.
  \item Schema, identity, and data conflicts.
  \item Incorrect or otherwise flawed data.
  \item Identification of a target schema and schema translation.
  \item Presentation of results.
\end{customitemize}

Data integration systems use various combinations of steps to cope with these challenges. Generally, these steps are:

\begin{enumerate}
  \item Schema mapping.
  \item Data source selection and data retrieval.
  \item Data transformations and schema translation.
  \item Duplicate detection (object identity resolution).
  \item Quality assessment.
  \item Data fusion and conflict resolution.
  \item Result loading or visualization.
\end{enumerate}

We aim to create a Linked Data integration framework covering all these steps in \term{\odcs}~\cite{wise2012odcs}. \odcs is a server application for integration and management of Linked Data. It accepts Linked Data as RDF,\footnote{\url{http://www.w3.org/TR/rdf-syntax/}} processes them in a customizable pipeline of data processing units and saves the result to a data store. Users are provided with integrated views on the processed data that are generated on demand by the data fusion component presented in this paper. This \term{partially materialized} approach gives us the flexibility of being able to add or change stored data or even schema mappings at any time without the need to re-run data fusion on the whole data store, thus giving us the advantages of gradual evolution in a \paygo manner. The whole process is designed so that the trust aspect is supported with provenance tracking and quality assessment of integrated data.

\subsection{Data Fusion}

The contribution of this paper covers the data fusion phase with conflict resolution and a conflict-aware quality assessment of fused data. We present new algorithms that are implemented in \odcs and are also available as a standalone tool \term{\odcsft}.\footnote{For the rest of this paper, we will use the name \term{\odcsft} to refer to both the implementation of data fusion in \odcs and in \odcsft.}

Data fusion is the step where actual data merging happens -- multiple records representing the same real-world object are combined into a single, consistent, and clean representation~\cite{bleiholder2010dissertation}. In order to fulfill this definition, we need to establish a representation of a record, purge uncertain or low-quality values, and resolve identity and other conflicts. Therefore we regard conflict resolution as a subtask of data fusion.

Conflicts in data emerge during the data fusion phase and can be classified as schema, identity, and data conflicts. Schema conflicts are caused by different source data schemata -- different attribute names, data representations (e.g., one or two attributes for name and surname), or semantics (e.g., units). Identity conflicts are a result of  different identifiers used for the same real-world objects. Finally, data conflicts occur when different conflicting values exist for an attribute of one object.

Conflict can be resolved on \term{entity} or \term{attribute} level by a \term{resolution function}. Resolution functions can be classified as \term{deciding functions}, which can only choose values from the input such as the maximum value, or \term{mediating functions}, which may produce new values such as average or sum~\cite{bleiholder2010dissertation}.

\begin{figure*}
    \centering
    \begin{subfigure}[b]{0.5\textwidth}
            \includegraphics[width=\textwidth]{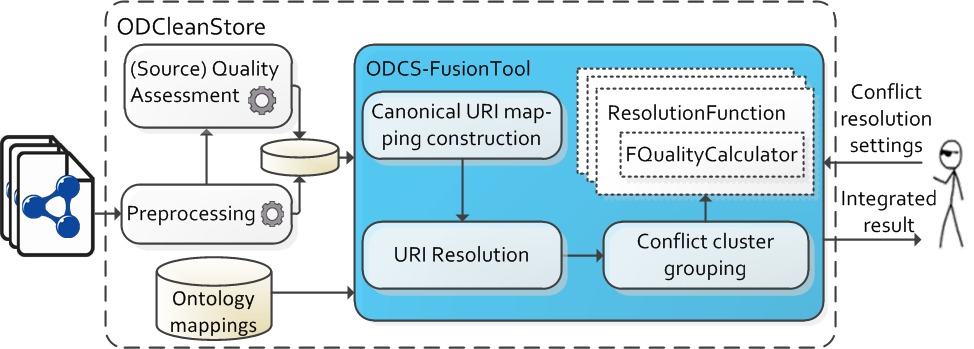}
            \caption{Data fusion in the \odcs framework}
            \label{fig:df-odcs}
    \end{subfigure}%
    \quad
    \begin{subfigure}[b]{0.47\textwidth}
            \includegraphics[width=\textwidth]{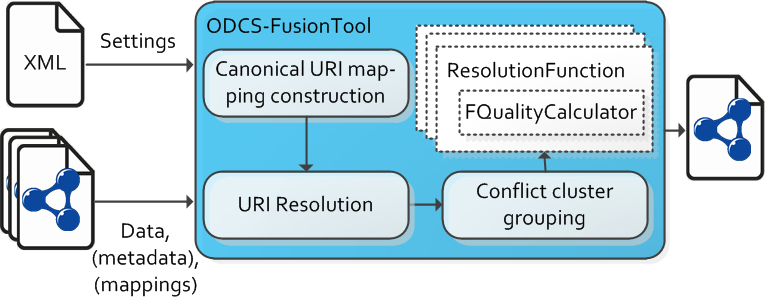}
            \caption{Data fusion with the standalone tool \odcsft}
            \label{fig:df-odcsft}
    \end{subfigure}
    \caption{High-level overview of the \odcsft implementation in the context of \odcs and when used independently.\\ \odcsft constructs a mapping to canonical URIs from input same-as links and ontology mappings in the first step which is used to resolve identifier and schema conflicts in the second step. The result is grouped to clusters of possibly conflicting triples where conflicts are resolved by a resolution function according to configuration. Resolution function also produces provenance and quality information about the integrated result.}
    \label{fig:df}
\end{figure*}

\medskip
The basic structure of the presented data fusion algorithm and its inputs are outlined in Figure~\ref{fig:df}. In the context of the \odcs framework (Figure~\ref{fig:df-odcs}), data fusion is executed at query time before query results are returned to the user. Application of the algorithm can be facilitated  by preprocessing steps such as data transformations or schema translation. The standalone implementation (Figure~\ref{fig:df-odcsft}) can be used for batch processing on raw RDF data stored in SPARQL endpoints or files. These can optionally also contain the metadata and mappings to be used in conflict resolution.

\subsection{Data Quality}
Measurement of information quality plays a fundamental role in data integration. Quality score may be the factor deciding which value to pick during conflict resolution~\cite{mendes2012sieve}; even if not quantified by a scoring function, quality indicators and metadata are commonly used in conflict resolution~\cite{bilke2005hummer, bizer2009wiqa, chawathe1994tsimmis, subrahmanian1995hermes}. These indicators may be both assessed automatically or entered manually by an expert user.

The quality of data produced \textit{from} a data integration process is also very important. It helps data consumers in decisions whether to trust a particular piece of information, or for overall quality assessment before further data processing. 

We introduce the concept of \term{\fq} as a measure of quality of integrated and fused data. 
We also propose a method of \fq assessment which has been implemented and evaluated in conjunction with our data fusion algorithm in \odcsft.
The assessment method is based on the conflict resolution context represented by conflicting statements, quality-related metadata, and user settings given to a conflict resolution function.
Compared to other quality assessment methods, leveraging the conflict resolution context allows data fusion to make decision even in the absence quality-related metadata using a voting-like approach.

\section{Related Work}
\label{sec:relatedWork}

\subsection{Linked Data Integration}
\label{sec:rwLdi}
In the young area of \ld, tools focus on individual aspects of \di, e.g., \oi or federated querying, but apart from \odcs only LDIF~\cite{schultz2011ldif} offers a complex \di solution. The same goes for data fusion and conflict resolution -- to the best of our knowledge, only Sieve~\cite{mendes2012sieve}, a part of LDIF, solves this task.

Linked Data Ingteration Framework (LDIF) is a framework producing homogenized views over heterogeneous data from diverse sources with a modular architecture. Modules include data retrieval, schema translation, \oi, quality assessment, and data fusion with conflict resolution (Sieve).

Sieve consists of two parts that run in sequence.
 First, Quality Assessment produces quality scores from user-selected metadata and configuration. The score calculation is based on concepts such as \term{assessment metric}, \term{data quality indicator}, or \term{scoring function}. In the second step, Data Fusion Module uses the quality scores in order to perform user-configurable conflict resolution. Sieve offers an extensible set of \term{fusion functions} used for resolution of conflicts.

\odcsft presented in this paper has a similar purpose as Sieve -- data fusion with conflict resolution. Due to a different intended use of the two, \odcsft has several unique features compared to Sieve. Sieve is used to fuse data as part of an ETL process realized by LDIF, i.e.~conflicts are resolved when data are loaded into a data store. \odcsft, on the other hand, is designed as an independent library which can be used either when loading data, or at query time, supporting a \paygo approach~\cite{paton2012pay}.

In Sieve, Quality Assessment produces scores through user-configured scoring functions and then Data Fusion is able to use them during conflict resolution. \odcsft doesn't include quality assessment (it is a separate module in \odcs) but uses arbitrary given RDF metadata as its input. More importantly, it produces quality scores of the \textit{fused} data, as opposed to source data. This way we can leverage the context information available in the fusion phase and reflect it in quality calculation, e.g. the quality may be different for a value confirmed by several sources and a value that is contradicted by other sources. In addition, it enables quality-based decisions even when source metadata are missing, as opposed to Sieve.

LDIF optionally outputs provenance metadata and results of Quality Assessment of source data. This is beyond the scope of \odcsft (metadata are given on demand in \odcs) but it outputs provenance of \textit{resolved} data so that data consumers can track where exactly the fused quads originated from. Another difference is that \odcsft can track quality to the quad level whereas LDIF tracks only to named graphs.

\subsection{Conflict Resolution}
Sieve, described in Section~\ref{sec:rwLdi}, is the only project that addresses conflict resolution for Linked Data. This section lists non-Linked Data integration systems that acknowledge conflicts and provide means for their resolution, sorted by descending year of publication. 

\textbf{Fusionplex}~\cite{motro2006fusionplex} integrates  heterogeneous information sources with resolution of factual inconsistencies. It uses metadata such as timeliness, accuracy or cost to compute \term{data utility}. Data can be filtered based on data utility or the actual values and fused by a selected fusion function (average, maximum, any, ...).
Data inconsistencies are handled in several steps: (1)  Tuples sharing the same key are grouped to \term{polytupes}. (2) Each polytupe is purged according to user preferences. (3) Values for each attribute are purged and fused to a single value, leaving a single tuple per polytupe.

\textbf{The Humboldt Merger}~\cite{bilke2005hummer} (HumMer)  allows ad-hoc, declarative fusion of conflicting data using an extension to SQL. It features schema matching, \oi, and data fusion/conflict resolution steps. All steps are executed ad-hoc at runtime in response to a user query. The user query is expressed in SQL extended with a \texttt{FUSE BY} statement and conflict resolution specifications. HumMer provides a range of conflict resolution functions including source preference, voting, most recent, or standard SQL aggregations. HumMer goes one step further then other systems and
optionally visualizes each intermediate step of data fusion with the possibility to interfere.

\textbf{DaQuinCIS}~\cite{scannapieco2004daquincis} has a module for resolution of data conflicts based on a custom $D^2Q$ data quality model. 

\textbf{FraQL}~\cite{sattler2000fraql} is a lightweight SQL extension for defining integrated object-relational schemata as well as formulating queries over them. FraQL recognizes several kinds of conflicts including both schema and data conflicts. Schema level conflicts can be resolved declaratively by attribute mappings or procedurally, data conflicts by user-defined reconciliation functions.

Mediation framework \textbf{AURORA}~\cite{yan1999aurora} proposes extension of SQL to support \term{conflict tolerant queries}, which produce conflict-free results in the presence of conflicts.
User specifies a fusion function for columns in the query result (e.g., \aggreg{Avg}, \aggreg{Max}, \aggreg{Any}, \aggreg{Discard}, user-defined). In addition, a tolerance strategy (one of \term{HighConfidence}, \term{RandomEvidence}, \term{PosibleAtAll}) may be given for evaluation of query conditions.



\textbf{$\mathrm{\mathbf{OO_{RA}}}$}~\cite{lim1998oora}
is an extended object-oriented data model and SQL extension to handle attribute-level conflicts where attribute resolution functions can be given. 

\textbf{Hermes}~\cite{subrahmanian1995hermes} integrates information from diverse sources (not limited to databases) employing declarative rule-based language for defining mediators in an extensible way with the ability to define  conflict resolution rules. The mediator may use predefined or custom strategies (e.g., \aggreg{Latest}, \aggreg{Max}).

\textbf{TSIMMIS}~\cite{chawathe1994tsimmis}
uses a custom \term{Object Exchange Model} based on quadruples (label, type, value, object-iid) extracted from the underlying sources. It recognizes possible inter-source duplicates and value conflicts, although only simple source preference is supported.

In \textbf{Multibase}~\cite{dayal1983multibase}, data describing the same type of entities are integrated using a principle of generalization (global classes generalize local ones) and SQL aggregation functions (\aggreg{Sum}, \aggreg{Avg}, \ldots) for inter-source conflicts.

\medskip

\odcsft is a more mature system in its modular architecture where quality computation and fusion functions are independent and pluggable. It also leverages OWL\footnote{\url{http://www.w3.org/TR/owl-features/}} to resolve schema and identity conflicts based on mappings. Finally, it is novel in that the quality calculation can leverage  the context of multiple data sources and it produces quality and provenance of the actual fused data rather then just source data. Several projects use declarative configuration by extending SQL. Extending SPARQL\footnote{\url{http://www.w3.org/TR/rdf-sparql-query/}} for such purpose is a topic for further research. Networked graphs~\cite{schenk2008networked} are a mechanism which can be used to declaratively integrate RDF data using SPARQL, although it doesn't support data conflict resolution directly.

\subsection{Conflict-Based Quality}
\label{sec:conflictBasedQuality}
A general framework for data fusion by Ronald Yager~\cite{yager2004framework} proposes a voting-like process to determine the best values. A total \term{support} is computed for each solution taking conflicts with other values into consideration. Using the approach described in Section~\ref{sec:withMethodology}, we have independently come to a solution  that  overlaps with Yager's framework.

In Yager's framework, a fusion engine has multiple sources on its input, each claiming a value $a_i$, and chooses the best fused solution $a$ based on user requirements, source credibilities, a proximity knowledge base and a knowledge of reasonableness.
A \term{support} of each source is computed for each possible $a$. The support is based on  proximity of $a$ and $a_i$ values weighted by credibility of sources. Support values are then combined together to produce a total support which may further be combined with a reasonableness value for each  $a_i$.


The results produced from \fq assessment proposed in this paper can be regarded as a special case of Yager's \term{support} with the following differences: \cite{yager2004framework} includes a domain-specific reasonableness in the calculation; \fq assessment instead goes one step further in considering confirmation of values by multiple sources.
\cite{yager2004framework} analyses requirements on  operators used in the support function and its properties. 
 These results apply to \fq as well.

\section{Data Fusion and Conflict Resolution Algorithm}
\label{sec:cr}

Let us recapitulate the main challenges of Linked Data fusion that need to be addressed:
\begin{enumerate}
  \item Different identifying URIs are used to represent the same real-world entities.
  \item Different schemata are used to describe data.
  \item Data conflicts emerge when RDF triples sharing the same subject and predicate have inconsistent values in place of the object.
\end{enumerate}

Context that the fusion algorithm can leverage consists of the data to be integrated, metadata
 and mappings between resource URIs and property URIs from the used schemata, and  conflict resolution settings. 
In this section, we introduce the necessary terminology, describe our proposed algorithm and analyze its time \& memory complexity.

We will demonstrate the algorithm on a simple example used throughout this section. Let us consider \exampleref{ex:data} with five statements\footnote{We use the TriG notation (\url{http://www.w3.org/TR/trig/})} from three sources that represent a label and the geographical longitude of the city of Berlin. We can see  different URIs are used to identify Berlin, properties from different vocabularies are used, and two conflicting values for each property are provided. Our task will be to fuse data about Berlin and provide a single best value for each property of Berlin, i.e. label and longitude.

\begin{examplelisting}{ex:data}{Sample data before data fusion}
\verb!GRAPH <http://dbpedia.org> {!\\
\verb!  db:Berlin  rdfs:label  "Berlin".!\\
\verb!  db:Berlin  geo:long    "13.399". }!\\
\verb!GRAPH <http://rdf.freebase.com> {!\\
\verb!  fb:en.berlin  rdfs:label  "Berlin". !\\
\verb!  fb:en.berlin  fbgeo:long  "13.383". }!\\
\verb!GRAPH <http://data.nytimes.com> {!\\
\verb!  nyt:N50987  skos:label  "Berlin (Germany)" }!\\
\end{examplelisting}

\subsection{Formalism}
\label{sec:crFormalism}
\begin{definition}[RDF nodes]
Let $U$, $B$ and $L$ be sets of all URI references, blank nodes and RDF literals, respectively. Sets $U$, $B$ and $L$ are pairwise disjoint. An RDF node is an element of their union $N=U\cup B\cup L$.
\end{definition}

\begin{definition}[RDF triple]
An RDF triple is a statement expressing that a resource has a property with a certain value. Formally, the set of all triples is
$$Triples = (U\cup B) \times U \times (U\cup B \cup L).$$
We denote the elements of each triple as \term{subject} (resource of interest), \term{predicate} (property), and \term{object} (value of the property), respectively.
\end{definition}

\begin{definition}[RDF graph, \ngraph]
A subset $G$ of $Triples$ can be represented as a directed labeled graph and we refer to it as an \term{RDF graph}.

A \term{\ngraph} is a pair $(G, n)$ where $G\subset Triples$ is an RDF graph and $n\in U$.
 We say that graph $G$ is \term{named} $n$.
\ngraph{}s cannot share blank nodes, i.e. blank nodes in triples from $(G_1, n_1)$ are distinct from those in $(G_2, n_2)$ for $n_1 \neq n_2$. 
\end{definition}

\begin{definition}[Quad]
Let \term{quad} denote a quadruple $(s,p,o,g)$ such that there is a triple $(s,p,o)$ in RDF graph $G$ named $g$.
The set of all quads is denoted: 
$$Quads = (U\cup B) \times U \times (U\cup B \cup L)\times U$$
\end{definition}

\begin{definition}
  Let $G$ be an RDF graph. We define functions $subjects(G)$, $predicates(G)$ and $objects(G)$ by the following formulas:
$  subjects(G) = \left\{s~|~(s,p,o)\in G \right\}$,
$  predicates(G) = \left\{p~|~(s,p,o)\in G \right\}$,
$  objects(G) = \left\{o~|~(s,p,o)\in G \right\}$
We define these functions on a set of quads $Q\subset Quads$ analogously and 
$graphs(Q) = \left\{g~|~(s,p,o,g)\in Q \right\}.$
\end{definition}

\label{sec:crTerminology}
The result of the conflict resolution algorithm implemented in \odcsft contains not only the RDF triples with conflicts resolved according to the given conflict resolution strategy but also:

\begin{enumerate}
  \item Names of the \ngraph{}s each triple was selected from or derived from.
  \item A quality value for each triple based on conflicting values and input metadata; quality is expressed as a value from an ordered space $C=[0;1]$. This quality (formalized in Definition~\ref{def:fqScore}) expresses how trustworthy the respective triple is with regards to other conflicting values and input metadata (such as data source quality).
\end{enumerate}

In order to convey this information in the result, we introduce \term{resolved quads} as output of the conflict resolution algorithm.

\begin{definition}[Resolved quad]\label{def:resolvedQuad}
  \term{Resolved quad} is a triple $(q,S,c)$ from the space of all result quads denoted $ResolvedQuads$ defined as
$$ ResolvedQuads = Quads \times \power{U} \times C,$$

where $q$ is a result quad, $S$ is the set of names of \ngraph{}s $q$ was selected or derived from ($\mathcal{P}$ denotes the power set) and $c$ is the quality. 
\end{definition}

\exampleref{ex:resolvedQuad} shows how a resolved quads produced for data from \exampleref{ex:data} may look like.

\begin{examplelisting}{ex:resolvedQuad}{Resolved quad}
\verb!((dbpedia:Berlin, rdfs:label, "Berlin", ex:1),!\\
\verb! {http://dbpedia.org,http://rdf.freebase.com},!\\
\verb! 0.71)!\\
\verb!((dbpedia:Berlin, geo:long, 13.391, ex:2),!\\
\verb! {http://dbpedia.org,http://rdf.freebase.com},!\\
\verb! 0.85)!
\end{examplelisting}

The proposed algorithm deals with conflicting property values, i.e. conflicts in place of quad objects.
This is sufficient to examine all triples and it naturally corresponds to attribute value resolution in traditional conflict resolution~\cite{bleiholder2010dissertation}.
Therefore quads which are in conflict must share the same subject and predicate. This motivates the following definition.

\begin{definition}[Object conflict cluster]
\label{def:objectConflictCluster}
  \term{Object conflict cluster} in a set of quads $Q\subseteq Quads$ is a  maximal subset $CC\subseteq Q$ such that $|subjects(CC)|=1$ and $|predicates(CC)|=1$.

In other words, all quads in a conflict cluster share the same subject and predicate. We will denote the conflict cluster with subject $s$ and predicate $p$ as $CC_{s,p}$.
\end{definition}

Conflict resolution functions in the context of relational databases typically operate on attribute values from the attribute's domain or on a set of tuples representing a database record. For Linked Data, the former approach would lead to loss of provenance information. Therefore whole quads must  be given to a conflict resolution function rather than conflicting (object) values.

Conflict resolution functions also need additional metadata and context information in order to have enough expressive power to implement all desired resolution strategies.
These metadata and context data can also be modeled in RDF as quads. Therefore, we use the following definition of a conflict resolution function.

\begin{definition}[Conflict resolution function]
\label{def:crFunction}
\term{Conflict resolution function} $f$ is a function
$$f: \power{Quads}\times \power{Quads} \rightarrow \power{ResolvedQuads}.$$
\term{Object conflict resolution function} $f'$ is a partial function 
  $$f': \power{Quads} \times \power{Quads} \rightarrow \power{ResolvedQuads}$$
such that the following holds for $f'(CC, M)$:
\begin{enumerate}
  \item If $CC$ is an object conflict cluster with subject $s$ and predicate $p$, then $f'(CC,M)$ is also a conflict cluster with the same subject and predicate.
  \item If $CC\neq \emptyset$ and $CC$ is not an object conflict cluster in $Quads$, then $f'(CC, M)$ is undefined.
\end{enumerate}
\end{definition}
The first argument $CC$ of an (object) conflict resolution function represents the (possibly) conflicting quads to be resolved. The second argument $M$ represents metadata and context information. 
The distinction between $M$ and $CC$ is necessary as quads in $M$ may be necessary to decide about resolution of quads in $CC$ but they are not meant to be resolved themselves.

Because our algorithm considers only object conflicts, we further refer to object conflict resolution functions simply as \term{resolution functions} and to object conflict clusters as \term{conflict clusters} with no ambiguity.

Resolution functions can be classified as \term{deciding}, which can only choose values from the input such as the maximum value,
 or \term{mediating}, which may produce new values such as average or sum~\cite{bleiholder2010dissertation}.

A comprehensive list of resolution functions relevant for Linked Data can be found in \ref{app:resolutionFunctions}.

\subsection{Input \& Output}
The presented conflict resolution algorithm takes the following inputs:
\begin{enumerate}
  \item Collection of $quads$ to be resolved.
  \item Metadata represented as a collection of $quads$.
  \item Mappings between URI resources.
  \item Conflict resolution policy which specifies the default conflict resolution strategy and optionally per-property resolution strategies.
\end{enumerate}

Sample of what values can be given for each of these items can be found in \exampleref{ex:input}.

\begin{examplelisting}{ex:input}{Sample data fusion algorithm inputs}
\begin{enumerate}
  \item Data: see \exampleref{ex:data}
  \item Metadata:\\ \texttt{<http://dbpedia.org> odcs:score "0.9"},\\
        \texttt{<http://rdf.freebase.com> odcs:score "0.8"},\\ 
        \texttt{<http://data.nytimes.com> odcs:score "0.8"}
  \item Mappings:\\
        \texttt{rdfs:label odcs:equivalent skos:prefLabel},\\
        \texttt{geo:long~~~odcs:equivalent fbgeo:long},\\
        \texttt{db:Berlin owl:sameAs fb:en.berlin},\\
        \texttt{db:Berlin owl:sameAs nyt:N50987}
  \item Conflict resolution policy:\\
        \texttt{<Resolution function="BEST">\\<Property id="rdfs:label"/></Resolution>}\\
        \texttt{<Resolution function="AVG">\\<Property id="geo:long"/></Resolution>}
\end{enumerate}
\end{examplelisting}

Metadata can be anything the chosen conflict resolution function needs to produce an appropriate output and assess its quality. For example, it can be the timestamp for each named graph occurring in the input when using the \aggreg{Latest} resolution function, and named graph quality or user preference used in fused data quality assessment (the quality assessment algorithm proposed in Section~\ref{sec:qa} can work even in the absence of such metadata).

The mappings between URI resources express both results of \oi and schema mappings. They are represented as RDF triples with \sameas\footnote{URI namespace prefixes used in this paper can be resolved with \url{http://prefix.cc/}} as their predicate -- e.g., a triple ($s$, \sameas, $o$) states that $s$ and $o$ represent the same thing and are equal for the purposes of conflict resolution. Sometimes it is inconvenient to use \sameas\ -- e.g., to map \texttt{dbp:father} and \texttt{dbp:mother} to target property \texttt{ex:parent} it is incorrect to state that these two are the same. A special property \texttt{odcs:equivalent} can be used in such cases instead.\footnote{ For simplicity, we will further refer to any mapping triples only as \term{\sameas links}.}

A \term{resolution strategy} defines how conflicts shall be resolved. Most importantly, it specifies the \term{resolution function} to be used. 

 The chosen resolution function may not be applicable for some values -- e.g., function calculating the numeric average is not applicable to string values. The \term{aggregation error strategy} gives the desired behavior in that case. Inappropriate values may be either discarded or propagated to the output unchanged.

 Other parameters affecting, e.g., fused data quality calculation can also be given.

The output of the algorithm is a collection of \term{resolved quads} (Definition~\ref{def:resolvedQuad}), i.e. quads resolved by their respective resolution function together with a quality value and source graph names for each quad.

\begin{figure*}[t]
    \centering
    \begin{subfigure}[t]{0.3\textwidth}
            \includegraphics[width=\textwidth]{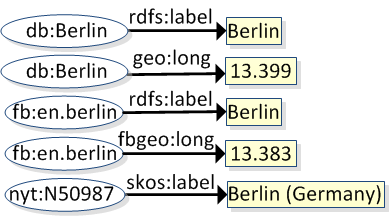}
            \subcaption{}
            \label{fig:fusion-example-before}
    \end{subfigure}%
    \quad
    \begin{subfigure}[t]{0.3\textwidth}
            \includegraphics[width=\textwidth]{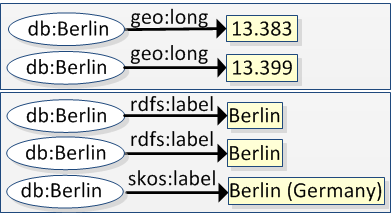}
            \subcaption{}
            \label{fig:fusion-example-during}
    \end{subfigure}%
    \quad
    \begin{subfigure}[t]{0.21\textwidth}
            \raisebox{0.6cm}{\includegraphics[width=\textwidth]{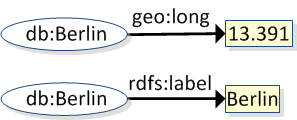}}
            \caption{}
            \label{fig:fusion-example-after}
    \end{subfigure}%
    \quad
    \caption{An example of RDF triples before, during, and after data fusion.}
    \label{fig:fusion-example}
\end{figure*}

\subsection{High-level Overview}\label{sec:crOverview}
The algorithm includes the following steps in order to deal with the challenges outlined earlier:

\begin{enumerate}
  \item Replace URI resources connected by a path of \sameas links 
    with a single URI (referred to as the \term{canonical URI}).
    \label{enum:hl-start}
  \item Remove duplicate identical quads.
  \item Group input quads into conflict clusters.
    \label{enum:hl-ccs}
  \item For each conflict cluster $CC_{s,p}$:
    \begin{enumerate}
      \item Choose a resolution function according to conflict resolution policy, predicate $p$ and \sameas mapping.
      \item Apply the resolution function. This includes resolution of conflicts, calculation of quality and provenance tracking of resolved quads. 
      \item Add the function's output to the result.
    \end{enumerate}
\end{enumerate}

Figure~\ref{fig:fusion-example} shows how  the algorithm is applied to our example. Figure~\ref{fig:fusion-example-during} shows the result after application of steps \ref{enum:hl-start}-\ref{enum:hl-ccs}: URIs are replaced with canonical variants and sorted triples are grouped into conflict clusters. The result in \ref{fig:fusion-example-after} is produced by selecting the highest-quality value for label and average for longitude.

\subsection{Algorithm Description}
Listing~\ref{algo:conflictResolution} is a formal description of the
presented 
conflict resolution algorithm. It uses functions $CanonicalMapping()$ to construct canonical URI mappings from input \sameas links (Listing~\ref{algo:canonicalMapping}), $UriResolution()$ to resolve identifier conflicts (Listing{}~\ref{algo:uriResolution}), groups data into conflict clusters, and finally applies an appropriate resolution function to each cluster.

\begin{customalgo}{algo:conflictResolution}{Conflict resolution algorithm}
  \REQUIRE Quads to be resolved $Q$, metadata quads $M$, 
     set of \sameas links $S\!A$, default conflict resolution strategy $defaultStrategy$, partial function $strategy(p)$ returning resolution strategy for predicate $p$.
  \ENSURE Collection of resolved quads.

  \STATE $result \gets \emptyset$
  \STATE $canonical\gets CanonicalMapping(S\!A)$ 
    \label{line:canonicalMappingCall}
  \STATE $Q\gets UriResolution(Q, canonical)$.
    \label{line:uriResolutionCall}
  \STATE Sort $Q$ lexicographically by quad subject, predicate, object and graph name.
    \label{line:crSort}
  \STATE Remove duplicate quads from $Q$.
    \label{line:crUnique}  
  \FORALL{$s,p$ such that $(s,p,o,g)\in Q$}
      \label{line:conflictClusterFor}
    \STATE 
      $CC_{s,p}\gets \left\{(s,p,o,g)~|~(s,p,o,g)\in Q\right\}$
      \label{line:conflictClusterCreate}
    \IF{$\exists p': canonical(p') = p$ \\ \qquad\AND $strategy(p')$ is defined}
        \label{line:crStrategyStart}
      \STATE $strategy\gets strategy(p')$.
    \ELSE
      \STATE $strategy \gets defaultStrategy$.
    \ENDIF 
    \STATE $f\gets$ resolution function according to $strategy$.
      \label{line:crResolutionFunctionSelection}
    \STATE $result \gets result \cup f(CC_{s,p}, M)$
  \ENDFOR \label{line:crEnd}
\end{customalgo}

\begin{customalgo}{algo:canonicalMapping}{Function $CanonicalMapping()$}
  \REQUIRE Set of \sameas links $S\!A$.
  \ENSURE Partial function $canonical:U\rightarrow U$.
  \STATE $canonical \gets \emptyset$.
  \STATE Create graph $H=(V,E)$ with \\vertices $V=subjects(S\!A)\cup objects(S\!A)$, \\and edges $E=\left\{\left\{s,o\right\}~|~(s,\mbox{\sameas},o)\in S\!A\right\}$.

  \STATE Find set of weakly connected components $\mathcal{C}$ in $H$.
  \FORALL{connected component $C=(V_C,E_C)\in\mathcal{C}$}
    \STATE Choose URI $c\in V_C$ as the canonical URI for $C$.
    \FORALL{$v\in V_C$}
      \STATE $canonical(v)\gets c$
    \ENDFOR
  \ENDFOR
  \RETURN $canonical$.
\end{customalgo}

\begin{customalgo}{algo:uriResolution}{Function $UriResolution()$}
  \REQUIRE Collection of quads to be resolved $Q$, canonical URI mapping $canonical:U\rightarrow U$.
  \ENSURE Collection of quads.
  
  \FORALL{input quad $q=(s,p,o,g)\in Q$}
    \IF{$s$ is a URI resource and $canonical(s)$ is defined}
      \label{line:uriResolutionReplaceSubjectStart}
      \STATE Replace $s$ in $q$ with $canonical(s)$.
    \ENDIF \label{line:uriResolutionReplaceSubjectEnd}
    \STATE Repeat steps~\ref{line:uriResolutionReplaceSubjectStart}-\ref{line:uriResolutionReplaceSubjectEnd} for predicate $p$ and object $o$.
  \ENDFOR
  \RETURN $Q$.
\end{customalgo}

\paragraph{Construction of Canonical URI Mapping}
The first step of the algorithm is the creation of canonical URI mappings (line~\ref{line:canonicalMappingCall} in Listing~\ref{algo:conflictResolution}). Data from multiple sources may use different URIs to represent the same concepts (identifier conflicts) and different predicate URIs for the same property (schema conflicts). We need to collapse URIs representing the same entity into a single \term{canonical URI}. Mapping from URI resources to their canonical URI is created from input \sameas links by function $CanonicalMapping()$.

The implementation of $CanonicalMapping(SA)$ is based on a disjoint-set data structure which keeps track of elements partitioned into a number of disjoint subsets, with operations $Find$ to determine the set an element belongs to, and $Union$ to join two subsets into one. 
 A forest-based implementation of disjoint-set can provide $\bigo(\alpha(n))$ amortized time per operation, where $\alpha$ is the inverse Ackermann function~\cite{tarjan1984dfu}. 
$CanonicalMapping()$ uses such data structure to partition URIs into sets.
For each link $(s,\sameas,o)$, $CanonicalMapping()$ calls $Union(s,o)$. Each subset then represents one weakly connected component of the \sameas links graph. One member of each set is chosen as the canonical URI.

\paragraph{URI Resolution}
The second step of the algorithm (line~\ref{line:uriResolutionCall} in Listing~\ref{algo:conflictResolution}, implementation in Listing~\ref{algo:uriResolution}) resolves identifier conflicts and schema conflicts according to given canonical URI mapping and prepares input for the application of a resolution function. This part is independent on the given conflict resolution policy.

\paragraph{Application of Resolution Function}
The rest of the conflict resolution algorithm in lines~\ref{line:conflictClusterFor}-\ref{line:crEnd} of Listing~\ref{algo:conflictResolution} groups quads into conflict clusters and applies a resolution function.

On line~\ref{line:crSort}, quads in $Q$ are sorted lexicographically by their subject, predicate, object and graph name. This allows efficient removal of duplicities on line~\ref{line:crUnique} (which may appear after the URI resolution step), and clustering of conflicting quads on lines~\ref{line:conflictClusterFor}-\ref{line:conflictClusterCreate} as quads from one object conflict cluster will be adjacent after sort.

The filtering procedure benefits from already resolved identifier conflicts.

The purpose of lines~\ref{line:crStrategyStart}-\ref{line:crResolutionFunctionSelection} is the selection of an appropriate resolution function. The user who entered resolution strategies for each property may have used a property URI different from what the algorithm selected as a canonical URI. Therefore we locate the appropriate resolution function using $canonical$ mapping.

The final step is the application of a resolution function. Implementations of resolution functions can be very diverse (see~\ref{app:resolutionFunctions}) but with respect to the expected output, they need to (1) produce result quads according to the resolution strategy they implement, (2) calculate quality of each result quad, and (3) keep track of \ngraph each result quad originates from.

Methods how to calculate  quality can also be diverse and depend on the resolution function and the task at hand. Our proposed quality assessment method can be found in Section~\ref{sec:qa}.

The architecture of \odcsft is flexible and both new resolution functions and quality assessment methods can be plugged in. 

\subsection{Time \& Memory Complexity}
In this section, we show that the time complexity of the algorithm is $\bigo(n\,\log n + l + T(n))$ and memory complexity is input size plus $\bigo(l)$ with few assumptions.

\begin{enumerate}
    \item Let $n=|Q|$ be the number of quads to be resolved.
    \item Let $l=|SA|$ be the number of \sameas links.
    \item Let $c=|\left\{(s,p);~(s,p,o,g)\in Q\right\}|$ be the number of conflict clusters.
    \item Let $c_i$ be the size of $i$-th conflict cluster.
    \item Let $T(x)$ be an upper bound of applied resolution functions' complexity.
\end{enumerate}

We assume the number of defined per-property resolution strategies is constant (independent on $n$ and $l$), and assume using a disjoint-set data structure with amortized time of operations $\alpha(x)$ as described above.

\paragraph{Time complexity}
There are several main operations affecting time complexity. The first is creation of canonical URI mapping in $\bigo(l\,\alpha(l))$; the second is translation to canonical URIs in $\bigo(n\,\alpha(l))$ (at most 3 lookups for each quad); third is sorting of quads in $\bigo(n\,\log n)$; the last one is application of resolution functions in total time $\bigo\left(\sum_{i=1}^{c} T(c_i)\right)$. Therefore the complexity is

$$\bigo\left(n \log n + (l+n)\,\alpha(l) + \sum_{i=1}^{c} T(c_i)\right).$$

We can simplify this formula with two other simple assumptions. Function $\alpha(l)$ grows very slowly and is less than five for all practical purposes. Let us further suppose that $T(x)=x^n$ ($T$ is polynomial). Using the binomial theorem, we can prove that the worst case occurs for $c=1$ and $c_1=n$ which gives us the worst-case time complexity
$\bigo(n\,\log n + l + T(n)). $

\paragraph{Memory complexity}
All operations on input quads $Q$ (URI resolution, sorting and removal of duplicates) can be executed in place. The only extra space is required for the disjoint-set representing the canonical URI mapping. Implementation as a tree with one node per URI requires $2l=\bigo(l)$ nodes.

\section{Fused Linked Data Quality Algorithm}
\label{sec:qa}

\term{Data quality} can be perceived as \quot{fitness of use} with respect to a particular task~\cite{juran1974quality}. Traditionally, quality of data is assessed before the data fusion phase based on factors such as timeliness or accuracy~\cite{wang1996accuracy, zaveriquality} relating to a single data source which prevents the fusion phase from consulting other sources. We propose a new approach that can leverage the context of multiple sources in quality assessment.

\subsection{\FQ}
\label{sec:dataQuality}
We introduce a new term \term{\fq} for the special case of fused data quality in order to distinguish the quality of source data (expressing quality of data per se) and quality of {fused} data (expressing quality in the context of other -- possibly conflicting -- sources).

The purpose of {\fq} is to (1)~act as one of the deciding factors during conflict resolution, (2)~help data consumers decide which data is worth using, and (3)~detect low quality data before further processing.
These goals are further supported in our solution by accompanying fused data with provenance metadata indicating where a particular value came from and thus supporting verifiability of information.

\begin{definition}[\FQ score]\label{def:fqScore}
  \term{\fq score} (or  \term{\fq} for short) is a number from interval $[0;1]$  expressing the quality 
  of a value after data fusion with respect to other conflicting values, content of the value, provenance of the value, and quality-related metadata. 
\end{definition}

The definition is intended for quality in the data integration context. Taking other conflicting values into consideration is what distinguishes it from how quality is regarded in this context, where only the rest of the factors are commonly used~\cite{zaveriquality}. 

For the purposes of Linked Data integration, we can replace the term \textit{value} in the definition with \textit{triple} or \textit{statement}. \fq is meant to be a factor used for conflict resolution decisions and also as a lead for data consumer's trust decisions about different values. For this reason, it is necessary to assess \fq at the statement or value level (rather than with lower granularity).

The range of \fq needs to be from an ordered space so that values can be easily compared. We define it as the interval $[0;1]$ where $0$ means no confidence and $1$ maximal confidence. This is in accordance with other related work where either the $[0;1]$ interval is used~\cite{mendes2012sieve, yager2004framework}, or simple binary decisions are made~\cite{bizer2009wiqa}. Range $[-1;1]$ has been proposed for the related concept of \term{trust}~\cite{hartig2008trustworthiness}, which is different from quality, however.

As stated in Section~\ref{sec:crTerminology}, the resolution of RDF statements rather than property values is convenient for \ld. Since we focus on resolution of RDF triple objects, we do not further distinguish between \fq of a resolved statement and its object.

\subsection{\FQ Assessment}
\label{sec:withMethodology}

We propose an  \fq assessment method based on three factors:
\begin{enumerate}
  \item Quality of data sources.
  \item Data conflicts.
  \item Confirmation of values by multiple sources.
\end{enumerate}
The exact steps of quality assessment may be customized for needs of each resolution function.

Quality of  source data is modelled on the \ngraph level -- \ngraph{}s are  convenient for attaching metadata to a set of triples and are commonly used in quality assessment~\cite{bizer2009wiqa, carroll2005namedgraphs, mendes2012sieve}. 
For a statement from a \ngraph $(G,g)$, we interpret the quality score associated with $(G,g)$ as the initial statement's quality.

Assessment of source \ngraph{}s' quality is \emph{not} part of our assessment algorithm. Quality dimensions can be described by various indicators~\cite{wang1996accuracy, zaveriquality} which need to be converted to a single quality score for our purposes. This task is beyond the scope of this paper and quality scores are simply an input value. A dedicated Quality Assessment module handles this task in \odcs~\cite{compsac2012aggregation}; other solutions include~\cite{flemming2010quality, hartig2008trustworthiness, hartig2009qa, mendes2012sieve}.

\paragraph{Methodology}
We took the following steps when designing our  assessment method.

First, we formulated objectives of \fq, as outlined in Section~\ref{sec:dataQuality}: act as a deciding factor in conflict resolution, help data consumers in decisions, and detect low data quality. Another objective was to enrich existing quality assessment methods with the ability to leverage the conflict resolution context.

As the next step, we identified information that can be used during assessment. Authors in~\cite{bizer2009wiqa} classify  quality indicators as based on information content itself, on metadata about the origin or on user ratings. The information content is represented by the conflicting statements, origin metadata and potential user ratings are modelled on the \ngraph level. They should be converted to a single score. Existing quality assessment tools can be utilized and the result quality score is given among metadata for the resolution function.

 Next, we collected several real-world cases for data integration~\cite{compsac2012aggregation}. The objectives and use cases lead us to several requirements on the assessment method summarized in Section~\ref{sec:qualityRequirements}.

We came up with several functions that attempted to satisfy the requirements. We chose one of the functions, which emerged as a simple and natural solution of the requirements, implemented it and proved its feasibility on evaluation in \odcs.

\paragraph{Terminology}
Let us pick up on the formalism introduced in Section~\ref{sec:crFormalism}:

\begin{definition}[\FQ function $q$]
  Let $q$ be a function
  $q:N\times \power{U}\times \power{Quads} \times\power{Quads}\rightarrow[0;1].$

  We interpret the value $q(\mathbf{v}, S, CC, M)$ as the \fq of value $\mathbf{v}$ (or resolved quad with $\mathbf{v}$ as its object, respectively).
  The value whose quality we calculate is represented by $\mathbf{v}\in N$. $S\subseteq U$ is the set of names of graphs which state this value (for deciding resolution functions\footnote{See Definition~\ref{def:crFunction}.}) or  which $\mathbf{v}$ was derived from (for mediating functions\footnotemark[\thefootnote]) such that $S\neq\emptyset$ and $S\subseteq graphs(CC)$. We refer to $S$ as  \term{source \ngraph{}s}. $CC$ is the set of conflicting statements (\term{conflict cluster}) and $M$ the set of metadata quads -- this is the same as for a resolution function.
\end{definition}



\begin{definition}[Distance measure]
\label{def:distanceMeasure}
  Distance measure is a function
  $d:N\times N\rightarrow [0;1]$
  such that $d$ is symmetric and $d(n,n)=0$. Value
  $d(n_1, n_2)$ is interpreted as the distance or difference of values represented by RDF nodes $n_1$ and $n_2$ where $0$ means most similar and $1$ means completely different.
\end{definition}

\begin{definition}[Graph quality score]
\label{def:graphQualityScore}
  Let $s:U\rightarrow [0;1]$ be a partial function which assigns quality score $s(g)$ to a \ngraph $(G,g)$.

The value of $s(g)$ is supposed to be either precomputed and given in metadata $M$ or it should be possible to calculate from $M$, therefore it doesn't need to be an explicit argument of the \fq function $q$.

  We will also use $\bar s:\power{U}\rightarrow [0;1]$ to denote a function which aggregates scores of several \ngraph{}s based on $s(g)$.

\end{definition}

\subsection{Requirements on \FQ Assessment}
\label{sec:qualityRequirements}

This section summarizes requirements on an \fq function. We have identified mathematical properties that should be satisfied and three main quality factors: (1) quality scores of source \ngraph{}s, (2) value conflicts and (3) confirmation by multiple sources.

\begin{enumerate}
  \item \label{item:requirementsMathStart} Function $q$ doesn't depend on the order of elements in $S$ or $CC$ (symmetry).
  \item Function $q$ is increasing with $\bar s(S)$ (monotonicity).
 \begin{align*}
\bar s(S_1)\geq \bar s(S_2)&~\Rightarrow \\ &q(\mathbf{v},S_1,CC,M)\geq q(\mathbf{v},S_2,CC,M)    
  \end{align*}

  \item $\bar s(S)$ is a function of $s(g_i)$, $g_i\in S$ and monotonous in its arguments ($\bar s$ monotonicity).
 \begin{align*}
 (S=&\left\{g_1,\ldots,g_n\right\}, S'=\left\{g_1',\ldots, g_n'\right\},\\ &\forall i=1,\ldots,n:s(g_i)\geq s(g_i'))\Rightarrow \bar s(S)\geq \bar s(S')
 \end{align*}
  \item Function $q$ decreases with increasing distance from other conflicting values $objects(CC)$. Significance of such difference should be weighted by the respective graph quality scores.
 \begin{align*}
(\forall v_i\in & objects(CC): d(\mathbf{v},v_i)\geq d(\mathbf{v'},v_i) )~\Rightarrow \\ & q(\mathbf{v},S,CC,M)\leq q(\mathbf{v'},S,CC,M)
\end{align*}
  \item \label{item:requirementsMathEnd} If $objects(CC)=\left\{\mathbf{v}\right\}$, then $q(\mathbf{v},S,CC,M)\geq q(\mathbf{v'},S,CC,M)$ for all $v'\neq v$ (idempotency).
  \item \label{item:requirementsIntuitiveStart}\label{item:requirementsAndSource} If there are no conflicts, i.e. $objects(CC)=\left\{\mathbf{v}\right\}$, then $q=\bar s(S)$.
  \item Sources with zero quality score should not affect $q$.
  \item If $n$ sources with the maximum score claim a value completely different from $\mathbf{v}$, it should decrease value of $q$ approximately $n$ times.
    \label{item:requirementsNTimes}
  \item If several sources agree on the same value $\mathbf{v}$, it increases the value of $q(\mathbf{v},S,CC,M)$.
    \label{item:requirementsIntuitiveEnd}
    \label{item:requirementsAgree}
\end{enumerate}

 Requirements~\ref{item:requirementsMathStart}-\ref{item:requirementsMathEnd} constraint mathematical properties of $q$ while requirements~\ref{item:requirementsIntuitiveStart}-\ref{item:requirementsIntuitiveEnd} were introduced so that  $q$  produces intuitive results. Our requirements  overlap with framework proposed in~\cite{yager2004framework} where a more detailed analysis of resulting properties of $q$ can be found. See also related work in Section~\ref{sec:conflictBasedQuality}.

\subsection{\fq Assessment Algorithm Description}
\label{sec:qualityAlgorithm}
In this section, we describe an \fq assessment method which complies with the requirements outlined in Section~\ref{sec:qualityRequirements}.

The assessment is called when a resolution function is executed. Therefore the details of the general assessment algorithm may be customized for the function -- we use a different version for mediating and deciding resolution functions in \odcsft, for instance.

The assessment algorithm is described in Listing~\ref{algo:qualityAssessment}, and details for each quality factor discussed below.

\begin{customalgo}{algo:qualityAssessment}{\fq assessment algorithm}
  \REQUIRE Assessed value $\mathbf{v}$, source graph names (for $\mathbf{v}$) S, collection of conflicting statements $CC = \left\{(s_1,p_1,o_1,g_1),\ldots,(s_n,p_n,o_n,g_n)\right\}$, metadata $M$.\\
    The algorithm is also parametrized with a distance measure $d$, graph quality score functions $s$ and $\bar s$, boolean constants $ConsiderConflicts$, $ConsiderConfirmation$ and a positive numeric constant $AgreeCoefficient$.
  
  \ENSURE \fq of value $\mathbf{v}$.
  \STATE $q\gets \bar s (S)$ \quad \COMMENT{Factor 1: Source quality}
    \label{step:qFactor1}
  \medskip
  \IF[Factor 2: Conflicting values]{$ConsiderConflicts$}
    \label{step:qFactor2Start}
    \STATE $\displaystyle conflictFactor\gets   1-\frac{\sum_{i=1}^n s(g_i)\cdot d(\mathbf{v},o_i)}{\sum_{i=1}^{n}s(g_i)}$
      \label{step:qFactor2coef}
    \STATE $q\gets q \cdot conflictFactor$
            \label{step:qFactor2and}
  \ENDIF \label{step:qFactor2End}
  \smallskip
  \IF[Factor 3: Confirmation by multiple sources]{$ConsiderSupport$}
    \label{step:qFactor3Start}
    \STATE $support = \left\{g_i~|~o_i = \mathbf{v}\right\}$
    \IF{$|support|>0$}
      \STATE $ supportFactor \gets \hfil\break \displaystyle\frac{ \left(\sum_{g_i\in support} s(g_i) - \max_{g_i\in support} s(g_i)\right)}{ AgreeCoefficient}$
        \vspace{5pt}
      \STATE $supportFactor = \min(supportFactor, 1)$
      \STATE $q\gets q + (1-q)\cdot supportFactor$
    \ENDIF
  \ENDIF \label{step:qFactor3End}
  \RETURN $q$.
\end{customalgo}

\subsubsection{Factor 1: Source Quality}
The first step is to use quality of sources (line~\ref{step:qFactor1}). Sources are represented by a set $S$ of URIs of \ngraph{}s value $\mathbf{v}$ originated from. Value of this parameter is resolution function dependent and supplied by the function. \odcsft uses  $S=\left\{g~|~(s,p,\mathbf{v},g)\in CC\right\}$ for deciding resolution functions (graphs that actually contain $\mathbf{v}$), and all graphs $S=\left\{g~|~(s,p,o,g)\in CC\right\}$ for mediating functions. When using the AVG (average) resolution function, for instance, all values contributed to the average indeed and should be considered as sources.

Function $\bar s$ computing total score from $s(g_i)$, $g_i\in S$ is also resolution-specific. $\bar s$ must be monotonous and symmetric, which is satisfied, e.g., by minimum, maximum, or average. \odcsft uses maximum for deciding functions and average for mediating functions.

\smallskip
We can continue with \exampleref{ex:input}: In \odcs, function $s(g_i)$ would return the value of \texttt{odcs:score} for $g_i$ in metadata $M$. Therefore, the value of $q$ after step one would be $q=0.9$ (maximum for \texttt{dbpedia.org} and \texttt{rdf.freebase.com}) for label $\mathbf{v}=$\texttt{"Berlin"}, and $q=0.85$ (average) for average longitude $\mathbf{v}=19.391$.

\subsubsection{Factor 2: Conflicting Values}
\label{sec:qaFactor2}
The second step (lines~\ref{step:qFactor2Start}-\ref{step:qFactor2End}) looks at the actual data to be resolved and conflicts between them using a distance measure $d$. The more a value deviates from what others claim, the lower  quality it has, taking source quality scores into consideration.

This step is executed optionally since different values may be acceptable in some cases, especially for many-valued properties such as \texttt{rdf:type}. There is a special user-provided resolution parameter which controls execution of this step in \odcsft at the global or per-property level.

\paragraph{Conflict Factor}
Line~\ref{step:qFactor2coef} calculates an average of value distances weighted by the respective quality scores. This can be interpreted as each source voting for value $\mathbf{v}$, its vote proportional to the similarity of $o_i$ claimed by the source to $\mathbf{v}$ and the vote weighted by source quality score. 

Multiplication in expression $s(g_i)\cdot d(\mathbf{v},o_i)$ on line~\ref{step:qFactor2coef} and on line~\ref{step:qFactor2and} could be replaced by any \term{and}-like operator. An \term{and}-like operator is  a binary operator $\wedge$ such that $\wedge$ is monotonous, $0\wedge x = x\wedge 0 = 0$ and $1\wedge x = x\wedge 1 = x$. This is satisfied, e.g., by multiplication or minimum.
We chose multiplication on line~\ref{step:qFactor2coef} because a weighted average is easily comprehensible and interpretable as \quot{voting}.
Multiplication on line~\ref{step:qFactor2and} was chosen over minimum to satisfy requirement~\ref{item:requirementsNTimes} and for the reason demonstrated by the following example. If we have a source with quality score $s(g_i)=0.5$ and $conflictFactor=0.5$, then we can intuitively expect that the result quality should be strictly less than when $s(g_i)=0.5$ and $conflictFactor = 1$ (or the other way round). Taking minimum wouldn't be able to distinguish these cases.


\paragraph{Distance Measure}
Function $d$ should be customized to the type of values it is applied to. Functions commonly employed in \oi are best suited for literals, e.g., Levenshtein or Jaro-Winkler distance for strings~\cite{schallehn2004similarity, volz2009silk} or  $\min(|v_1-v_2|/max, 1)$ for ordinal values ($max$ being a parameter)~\cite{schallehn2004similarity, yager2004framework}.
A domain-specific measure can be utilized for URI resources, e.g., value based on color similarity can be returned for resources representing colors like \texttt{ex:black}, \texttt{ex:gray}, etc. \odcsft uses Levenshtein for strings,  $\min(|2(v_1-v_2)/(v_1+v_2)|,1)$ for numbers (distance normalized by average), $\min(|v_1-v_2|/max, 1)$ for dates and inequality indicator (0 for equal values, 1 for others) otherwise.

\smallskip
In our working example, quality $q$ would decrease for label \texttt{"Berlin"} due to its difference from label \texttt{"Berlin (Germany)"}, and quality $q$ for longitude would decrease due to the difference between  average longitude and source longitude values.

\todo{interpretace support pokud je hodnota kompatibilní s daným zdrojem *a* zdroj je důvěryhodný?}

\subsubsection{Factor 3: Confirmation by Multiple Sources}
This last step (lines~\ref{step:qFactor3Start}-\ref{step:qFactor3End}) is motivated by requirement~\ref{item:requirementsAgree}. If several sources agree exactly on value $\mathbf{v}$, than it should have a higher \fq than any of the sources alone. Let us imagine three sources $g_1$, $g_2$, and $g_3$ each having source quality score 0.5 and each claiming the same value $\mathbf{v}$. Intuitively, one can  trust the value $\mathbf{v}$ more than if there was only one source supporting it. In this example, the resulting \fq would be $q=0.5$ without this factor, while it yields $q=0.75$ when $AgreeCoefficient$ is 2.


This step is parametrized with a constant $AgreeCoefficient\in\mathbb{R}^{+}$. The idea is that the \fq of $\mathbf{v}$ should increase linearly with the sum of source quality scores $\sum_{g_i\in support}s(g_i)$ so that
\begin{customitemize}
  \item if $|support| = 1$, then $q$ is unchanged,
  \item the maximum \fq of 1 is reached when $AgreeCoefficient\;+\;1$ sources agree on $\mathbf{v}$; more formally when $$AgreeCoefficient+\max_{g_i\in support} s(g_i)=\sum_{g_i\in support}s(g_i).$$
\end{customitemize}

This step is also optional because it doesn't make sense for resolution functions such as \aggreg{Avg} or \aggreg{Concat}\footnote{See \ref{app:resolutionFunctions}} where a value match would be coincidental and shouldn't increase the \fq.

\smallskip
In our working example, value of $q$ for label \texttt{"Berlin"} would increase because two sources agree on the value, but $q$ for longitude wouldn't change because factor 3 is not used for the \aggreg{Avg} resolution function.

\subsection{Complexity \& Relation to Resolution Functions}

\begin{figure*}[bt]
\centering
\setlength{\tabcolsep}{0.5ex}
\begin{tabularx}{\textwidth}{|c|c|c|>{\raggedright\ttfamily}X|c|@{}p{0cm}@{}}
  \hline
  Type & $Consider$- & $Consider$-  & \rmfamily \centering Resolution function & Time  & \\
       & $Conflicts$ & $Support$    &                               & complexity & \\
  \hline
  Deciding & yes & yes & All, Best, TopN, Threshold & $\bigo(n^2)$ & \\
  \cline{4-5}
  &  & & Any, BestSource, Filter, Longest, Max, Min, None, Shortest, Vote, WeightedVote & $\bigo(n)$ & \\
  \cline{4-5}
  &  & & MaxSourceMetadata, MinSourceMetadata & $\bigo(n \log |M|)$ & \\
  \hline
  Mediating & yes & no & Avg, Median & $\bigo(n)$ & \\
  \hline
  Mediating & no & no & Concat, Sum & $\bigo(n)$ & \\ 
  \hline
\end{tabularx}
\caption{Complexity and \fq settings of resolution functions in \odcs. See \ref{app:resolutionFunctions} for description of functions.}
\label{tbl:resolutionComplexity}
\end{figure*}

The time complexity of the \fq assessment algorithm is linear in the number or quads in the given conflict cluster. This determines the time complexity of resolution functions which are summarized in Figure~\ref{tbl:resolutionComplexity}.

We assume that evaluation of $s(g)$ and $d(x,y)$ is $\bigo(1)$, and  $\bar s(S)$ is $\bigo(|S|)$ (true in \odcs). Factor 1 is executed in $\bigo(|S|)=\bigo(n)$ because $S\subseteq graphs(CC)$. Factor 2 has a loop with $n$ iterations in constant time. Factor 3 can be implemented in one pass over the conflict factor. Therefore the total complexity is $\bigo(n)$.

Figure~\ref{tbl:resolutionComplexity} gives an overview of default resolution functions in \odcsft, their time complexity and quality assessment settings. These results further assume that metadata $M$ can be queried for quads with a known subject and predicate in $\bigo(\log |M|)$, conflict cluster is sorted from the conflict resolution algorithm (Listing~\ref{algo:conflictResolution}), and resolution functions are applicable to all values. \todo{see aggr error}
The complexity is quadratic for resolution functions \aggreg{All}, \aggreg{Best}, \aggreg{TopN} and  \aggreg{Threshold} because \fq must be evaluated for all quads in a conflict cluster, each needing an $\bigo(n)$ time. Other resolution functions evaluate quality assessment only for the one resolved quad they return. 

\subsection{Discussion}
The proposed \fq assessment algorithm covers the goals outlined above and gives results  easy to interpret. The assessment algorithm is independent on the quality assessment of data sources which may be task-specific and is beyond the scope of this paper.
A downside of the \fq assessment is that considering context represented by the conflict cluster $CC$ requires time $\bigo(|CC|)$ which results in a quadratic complexity of some resolution functions.

The presented assessment algorithm provides a basic framework for quality assessment in a data integration environment, taking data conflicts into consideration. Information quality is strongly task-dependent~\cite{bizer2009wiqa}, however. Data consumers have different requirements for a medical application and  a simple music database. In addition, quality assessment needs to be tailored for the resolution function it is used with (Figure~\ref{tbl:resolutionComplexity}).

For this reason, quality calculation is not a part of the conflict resolution algorithm but it is a part of resolution functions. Custom resolution functions with custom quality formulas may be plugged-in and the formula suggested here may serve as a basis for custom-tailored methods.

\section{Evaluation}
\label{sec:evaluation}
We have implemented and evaluated the proposed data fusion and quality assessment algorithms in the \odcs framework and also in a standalone tool \odcsft. The experience from more than a year of practical use has proven their feasibility and usefulness. In this section, we present our evaluation on publicly available Linked Data used in the OAEI evaluation campaign. We investigate both performance and effectiveness of the presented algorithms. 

For the evaluation, we chose the dataset used for the Instance Matching challenge of OAEI 2011\footnote{http://oaei.ontologymatching.org/2011/instance/} including data about locations from The New York Times (NYT), Freebase, DBPedia, and Geonames datasets because NYT contains high-quality curated links among these datasets and therefore the quality of links doesn't impair the results. The data were supplemented with ontology mappings available for Geonames and mappings handcrafted for the most frequent properties and classes in the datasets.
The dataset contained information about 3840 locations from NYT described by 2 million triples. 

The prepared data were integrated with three resolution functions: \aggreg{All}, \aggreg{Any} and  \aggreg{Best}.
Figure~\ref{tbl:evaluationStats} summarizes dataset statistics before and after integration.

\begin{figure*}
\centering
\begin{tabular}{|r||r|r|r|r|r||r|r|}
\hline
& \centering NYT & \centering DBPedia & \centering Freebase  & \centering Geonames & {\centering All data} & {\centering  \bfseries \aggreg{All}} & {\centering \bfseries  \aggreg{Best}/\aggreg{Any}}  \\
\hline
Total triples & 117,623 &  260,426 & 1,567,454 & 55,535 & 2,005,058 & 1,461,449 & 461,449 \\
\hline
\sameas links & 17,572 & 41,702  &0 & 26 & 67,340 & 67,340 & 67,340  \\
\hline
All subjects & 5,620 &2,100 & 1,917 & 3,544 & 11,481 & 6,624 & 6,624  \\
\hline
Unique subjects & 3,832 & 2,083 & 1,917 & 3,544 & 6,624 & 6,624 &  6,624  \\
\hline
Conflict clusters & 55,894 & 168,188 & 268,289 & 43,296 & 456,957 & 461,449 & 461,449  \\
\hline
Average cluster size & 2.10 & 1.55 & 5.84 & 1.28 & 4.39  & 4.11 & 1 \\
\hline
Unique predicates & 36 & 2,870 & 1,323 & 33 & 4,221  & 4,221 & 4,221 \\ 
\hline
\end{tabular}
\caption[Information about source evaluation datasets]{Evaluation dataset statistics before and after integration. \quot{All data} include additional \sameas mappings.}
\label{tbl:evaluationStats}
\end{figure*}

\subsection{Completeness, Conciseness, Consistency}
Increasing \term{completeness}, \term{conciseness} and \term{correctness} are three broad goals of data integration~\cite{dong2009fusion}. Correctness expresses how much the data conform to the real world while completeness and conciseness are in a way analogous to recall and precision in information retrieval. 

Completeness measures the amount of data relative to all available data in the given domain and is achieved by adding more data. We distinguish extensional and intensional completeness defined by the following formulas:
\begin{align*}
    \mbox{\textit{ext. completeness}} &= \frac{\ |~\mbox{\textit{unique objects in dataset}}~|\ }{\ |\mbox{\textit{unique objects in universe}}|\ }\\[1mm]
    \mbox{\textit{int. completeness}} &= \frac{\ |\mbox{\textit{unique attributes in dataset}}|\ }{\ |~\mbox{\textit{unique available attributes}}~|\ }
\end{align*}

Conciseness measures the uniqueness of object representations. Conciseness is increased when redundant data are removed.
 Again, we recognize extensional and intensional conciseness: 
\begin{align*}
    \mbox{\textit{ext. conciseness}} &= \frac{\ |\mbox{\textit{unique objects in dataset}}|\ }{\ |\mbox{\textit{all  objects in dataset}}|\ }\\[1mm]
    \mbox{\textit{int. conciseness}} &= \frac{\ |\mbox{\textit{unique attributes in dataset}}|\ }{\ |\mbox{\textit{all attributes in dataset}}|\ }
\end{align*}

\term{Consistency}, as defined in~\cite{mendes2012sieve}, is related to correctness but easier to measure. It measures conflicts in the datasets regardless of correct real world values.
$$
  \mbox{\textit{consistency}} = \frac{\ |\mbox{\textit{conflict clusters without conflicts}}|\ }{\ |\mbox{\textit{all conflict clusters}}|\ }
$$

For the purposes of our evaluation, we identify the number of objects in the universe with the number of unique entities in the source datasets combined; we also do the same with attributes. Consistency is measured as the ratio of conflict clusters without conflict (i.e. with a single unique object) to the total number of conflict clusters \emph{excluding} clusters for manyvalued properties. These properties have cardinality larger than one and their different values doesn't affect consistency. 

\begin{figure*}
\centering
\begin{tabular}{|r||r|r|r|r|r||r|r|}
\hline
& \centering NYT & \centering DBPedia & \centering Freebase  & \centering Geonames & {\centering All data} & {\centering  \bfseries \aggreg{All}} & {\centering \bfseries  \aggreg{Best}/\aggreg{Any}} \\
\hline
Ext. completeness & 57.9\% & 31.5\% & 29.0\% & 53.5\% & 100.0\% & 100.0\% & 100.0\% \\
\hline
Int. completeness & 0.9\% & 68.0\% & 31.3\% & 0.8\% & 100.0\% & 100.0\% & 100.0\% \\
\hline
Ext. conciseness & 68.2\% & 99.2\% & 100.0\% & 100.0\% & 57.7\% & 100.0\% & 100.0\% \\
\hline
Int. conciseness & 100.0\% & 100.0\% & 100.0\% & 100.0\% & 97.3\% & 100.0\% & 100.0\% \\
\hline
Consistency & 90.9\% & 88.5\% & 73.4\% & 97.0\%  & 78.7\% & 78.9\% & 100.0\% \\
\hline
\end{tabular}
\caption[Completeness, Conciseness and Consistency of evaluated data sources]{Completeness, Conciseness and Consistency of the evaluated data sets relative to the result of integration using \aggreg{Best}.}
\label{tbl:evaluationImprovement}
\end{figure*}


\paragraph{Results}
Measurements for each of the source datasets and all data combined are listed in Figure~\ref{tbl:evaluationImprovement}. The listed numbers are relative to the  result integrated using the \aggreg{Best} resolution function -- values of completeness, conciseness and consistency are therefore 100\% after integration for all three resolution functions except for consistency of 78.9\% measured on data resolved by \aggreg{All}. 

We can see that completeness is raised by simply putting  data together. However, this results in decrease of conciseness and consistency which must be solved by the data fusion process. \odcsft raised extensional conciseness by 73\% and intentional conciseness by 3\%. Improvement in consistency depends on configuration of conflict resolution -- while returning all conflicting values has a minimal impact on consistency, choosing a single best value improves it by 27\% in our evaluation.

A similar experiment on a dataset of Brazilian municipalities was conducted for Sieve in~\cite{mendes2012sieve}. Their results apply to \odcsft as well because it can be configured to produce the same results 
 on the dataset.

\subsection{Performance}
The runtime of each integration run was measured on a server with four 3.00GHz CPUs and 16 GB of memory. Figure~\ref{tbl:evaluationRuntime} lists the results.
 The run times are median values over three runs.

\begin{figure}
\centering
\setlength{\tabcolsep}{1ex}
\begin{tabular}{|r||r|r|r|}
\hline
&  \centering\aggreg{All} & \centering\aggreg{Best} & {\centering\aggreg{Any}}  \\
\hline
Initialization & 0:01.0 & 0:00.9  & 0:01.0  \\
\hline 
Triple loading & 8:45.8 & 8:36.8 & 8:32.7  \\
\hline
Conflict Resolution & 0:44.2 & 0:38.7 & 0:10.8 \\
\hline
Total & 9:52.2 & 9:37.1 & 9:03.7 \\
\hline
\hline
\small CR throughput (triple/s) & 45,360 & 51,810 & 185,650 \\
\hline
\small CR throughput (CCs/s) & 10,334 & 11,807 & 42,310 \\
\hline
\small Total throughput (triple/s) & 771 & 791 & 863 \\
\hline
\end{tabular}
\caption[Run times of \odcsft on sample datasets]{Run times of on sample datasets (min:sec). \quot{CCs} stands for conflict clusters.}
\label{tbl:evaluationRuntime}
\end{figure}

Several facts can be observed from the results. One is that the initialization phase, which includes resolution of canonical URIs, is very fast. The resolution would scale well even for a  large number of \sameas links thanks to its nearly linear time complexity.

Most time is consumed by triple loading. This issue deserve more attention in future work. Note that this problem is visible only when processing data in large batches. This is not the case in \odcs where conflicts are resolved on demand for a single query.

Run times of the actual data fusion/conflict resolution part differ for \aggreg{All} and \aggreg{Best} resolution functions, and for resolution function \aggreg{Any}. This is according to expectations -- the former functions have quadratic time complexity in the size of a conflict cluster while the latter is linear (see Figure~\ref{tbl:resolutionComplexity}). The throughput is satisfactory and conflict resolution scales linearly for typical data -- even for large number of triples, the size of conflict clusters is typically limited by a reasonable constant. The average size of a conflict cluster was 4.39 on the test data, with maximum of 279 triples in a single cluster.

\subsubsection{Quality}
In this section, we look closer at \fq values produced by our algorithm. We selected several properties from resource description of the city of Berlin in the evaluation dataset. \fq calculation uses metadata about quality of sources. We added sample values expressing preference of data from DBPedia (with score 0.9) over other sources (with score 0.8):

\begin{verbatim}
<http://dbpedia.org>      odcs:score "0.9".
<http://rdf.freebase.com> odcs:score "0.8".
<http://sws.geonames.org> odcs:score "0.8".
<http://data.nytimes.com> odcs:score "0.8".
<http://example.com/err>  odcs:score "0.8".
\end{verbatim}

In order to demonstrate behavior in occurrence of errors, we also added an artificial source \quot{Err} claiming an incorrect value of geographical latitude of Berlin. Figure~\ref{tbl:evaluationQuality} lists values of the selected properties together with provenance information and \fq produced by \odcsft for the sample data. The resolution function used is \aggreg{All}.

\begin{figure*}
\setlength{\tabcolsep}{0.3ex}
\centering
\begin{tabular}{|>{\ttfamily}l|>{\ttfamily}l|l|l|}
\hline
\rmfamily Property & \rmfamily Value & \FQ & Sources    \\
\hline
\hline
{rdfs:label} & "Berlin" & 0.75992 & DBPedia, Freebase, Geonames\\
& "City\_of\_Berlin" & 0.27447 & Freebase \\
& "Berlin (Germany)" & 0.22126 & NYT \\
\hline
{geo:lat} & "52.5006" & 0.72418 & DBPedia \\
& "52.5167" & 0.64381 & NYT \\
& "52.5233" & 0.64380 & Freebase \\
& "52.52437" & 0.64380 & Geonames \\
& "13.4126" & 0.15610 & Err \\
\hline
geo:long & "13.3989" & 0.89957 & DBPedia \\
& "13.4" & 0.79965 & NYT \\
& "13.41053" & 0.79963 & Geonames \\
& "13.4127" & 0.79956 & Freebase \\
\hline
{dbprop:web} & <http://www.berlin.de...php> & 0.37739 & DBPedia,Freebase\\
 & {<http://berlin.unlike.net/>} & 0.11793 & DBPedia \\
 & <http://www.berlin.de> & 0.09275 & Freebase \\
 & \ldots &&\\
\hline
{rdf:type} & schema:City & 0.92000 & DBPedia,Freebase\\
& schema:Place & 0.90000 & DBPedia \\
& geonames:Feature & 0.80000 & Geonames \\
& \ldots &&\\
\hline
\end{tabular}
\caption{Sample \fq values for integrated resource description of Berlin}
\label{tbl:evaluationQuality}
\end{figure*}

If we look at \fq of labels, we see \texttt{"Berlin"} is clearly the best value. This is expected as three sources agree on it. \fq of other labels depends mostly on their similarity to \texttt{"Berlin"}.

The errorneous value of latitude has been clearly determined as low-quality because it is in conflict with other sources. Latitude from the preferred source (DBPedia) is ranked highest. The remaining  values are ranked according to how close they are to the (weighted) average of all values. The same holds for longitude.

The website suggested by two sources is ranked best while remaining values are ranked according to quality score of their source.

Values of \texttt{rdf:type} are special in that the cardinality parameter was set to \texttt{MANYVALUED} for them. This corresponds to $ConsiderConflicts=false$ in the \fq algorithm, therefore conflicts between values do not decrease the quality. One can see that \fq then depends only on score of the underlying source and support by multiple sources.

\smallskip
It is important to realize that the relative \fq for each value is more important than absolute values. The purpose of \fq was established to be (1) a deciding factor for conflict resolution, (2) a decision support for  users and (3) an indicator of low-quality data. This goal was indeed achieved on the sample data. Using the \texttt{Best} resolution function would really return the most reasonable value for each property, a user would be given a good lead on which website is worth visiting and the errorneous value of latitude can be detected.


\todo{evaluace na TED & spol - valuable hints - see FW}
\todo{mention Virtuoso in experiments}

\section{Specifics of Linked Data Integration}
\label{sec:specifics}
We conducted a thorough comparison of Linked Data integration compared to relational databases as part of our work on \odcsft. This section summarizes its basic specifics and open challenges.

\paragraph{Resource descriptions}
Data fusion deals with \term{records}~\cite{bleiholder2010dissertation}. While this term naturally matches tuples in relational databases, its equivalent in \ld is ambiguous. The most common approach is with \term{resource descriptions} composed of RDF triples having the given RDF resource as its subject. It is sufficient in most scenarios and enables straightforward application of traditional database data fusion techniques. Other options are possible, such as the inclusion of adjacent blank nodes~\cite{w3cbd}. Including triples having the resource of interest in place of object is also an option.

\paragraph{Blank nodes}
Blank nodes (RDF resources without an identifier) introduce additional complexity in \ld integration across several steps. One problem is with query \quot{round-tripping} which affects the data retrieval step. If a result of a SPARQL query contains a blank node \texttt{\_:b1} in an object and we want to retrieve the resource description of \texttt{\_:b1}, we cannot reference it in a subsequent query because \texttt{\_:b1} has no longer any relation to the source graph~\cite{mallea2011blank}.

 The lack of global identifiers makes it impossible for \oi to produce mappings between them.

 The data fusion step can interpret blank nodes as either autonomous entities, or structured properties. In the latter case, it is possible to treat the structured property components independently or together.

  Finally, our \fq assessment algorithm uses a distance measure to compare values -- comparing blank nodes as structured attributes is computationally difficult~\cite{carroll2003signing}.

\paragraph{Schemata}
The equivalent of schema in RDF is an ontology, typically expressed in OWL. While ontology can define restrictions, they are often not applied strictly~\cite{hogan2010weaving}.
Schemata are rather loosely used, and properties and classes need not be defined explicitly. This makes automation difficult (e.g., automatic generation of conflict resolution settings~\cite{volha2014sievelearner}). Another implication is that satisfaction of integrity constraints cannot be reliably used to detect low quality data. On the other hand, this approach is very flexible and target schema can evolve more easily. Schema mappings can also be richer (e.g., subproperty/subclasss taxonomies).

\paragraph{Query Execution} Implementations of the fusion step in relational databases are built on non-trivial join- and union-based operations~\cite{bleiholder2010dissertation} in order to align schema and join data from different sources. RDF simplifies this aspect to simply selecting relevant triples. In addition, we have the power of SPARQL and inferencing at hand.

\paragraph{Nulls}
The special value \dbnull has no equivalent in RDF other than simply a missing value. This makes it difficult to distinguish uncertainty and contradiction -- e.g., when one source contains a book's reviewer while other doesn't, the reviewer may be unknown (uncertainty) or the book wasn't reviewed (contradiction).

\section{Conclusion}
\label{sec:conclusion}
A wider adoption of \ld needs a critical amount of data, experience, and tools. This paper contributes to the arsenal of available tools with \odcsft, a data fusion and conflict resolution tool with quality assessment and provenance tracking. We also contribute to the data integration experience with a summary of \ld integration specifics and an overview of possible conflict resolution functions (\ref{app:resolutionFunctions}).
\todo{contribution by také mohly být experiences from evaluation}

Our implementation is based on new well-defined algorithms. One  algorithm fuses data with resolution of identifier, schema, and data conflicts. The other one computes fused data quality  for which we introduce a new term \term{\fq} -- it is novel in quality assessment of fused data rather than source data, and leveraging the context of multiple (possibly conflicting) sources. The combination of the two algorithms is mutually beneficial and enables data fusion to make quality-based decisions even in the absence of quality-related metadata.

\odcsft has been evaluated on real-world Open \ld. We achieved a distinct improvement of conciseness and consistency and gained valuable observation for future research.
Usefulness and feasibility has also been proven by more than a year of active usage in relation to the OpenData.cz initiative\footnote{\url{http://opendata.cz}} on processing of Czech public contracts data and EU procurement notices.\footnote{\url{http://isvzus.cz}, \url{ted.europa.eu}}

\subsection{Future work}
One area for further research are blank nodes. The proposed data fusion algorithm uses triples sharing the same subject as resource descriptions. An open question is how to effectively leverage inclusion of blank nodes in resource descriptions and decide between interpreting them as entities or structured attributes.

An interesting option is integration of conflict resolution with the SPARQL query language like other systems extend SQL \cite{bilke2005hummer, sattler2000fraql, yan1999aurora}. Currently, data need to be loaded beforehand and conflict resolution settings have no effect when querying the underlying RDF store.

Our experiments show data retrieval as the performance bottleneck. SPARQL queries executed on the underlying RDF store constituted up to 90\% of execution time. More efficient access to data, caching and parallelization should be investigated. Our algorithm is parallelizable in that once canonical URIs are resolved, resource descriptions are resolved independently.

\section*{References}
\bibliographystyle{elsart-num-sort}
\bibliography{bibliography}

\appendix
\section{List of Conflict Resolution Functions}
\label{app:resolutionFunctions}

A number of functions for resolution of conflicting values has been proposed in the literature \cite{bilke2005hummer, bleiholder2010dissertation, wise2012odcs, motro2006fusionplex,  subrahmanian1995hermes, yan1999aurora}. This appendix contains a comprehensive overview of functions relevant for \ld  proposed in the literature with a few new additions. Some systems also allow custom user-defined resolution functions~\cite{sattler2000fraql}.

\begin{description}
\setlength{\itemsep}{0em}
\small
  \item[All.] Returns all values.
  \item[Any.] Returns an arbitrary (non-\dbnull) value.
  \item[First, Last.] Returns the first or the last (non-\dbnull) value, respectively. Requires ordering of the values on input.
  \item[Random.] Returns a random (non-\dbnull) value. The chosen value differs among calls on the same input.
  \item[Certain.] If input values contain only one distinct (non-\dbnull) value, returns it. Otherwise returns \dbnull or empty output (depending on the underlying data model).
  \item[Best.] Returns the value with the highest data quality value. The quality measure is application-specific.
  \item[TopN.] Returns $n$ best values (see \aggreg{Best}). $n$ is a parameter.
  \item[Threshold.] Returns values with data quality higher then a given threshold. The threshold is given as a parameter.
  \item[BestSource.] Returns a value from the most preferred source. The preference of source may be explicit (given preferred order of sources) or based on an underlying data quality model. 
  \item[MaxSourceMetadata.] Returns a value from the source with a maximal source metadata value. The metadata value may be, e.g., timestamp of the source, access cost or a data quality indicator. The used type of source metadata is either given as a parameter or fixed.
  \item[MinSourceMetadata.] Returns a value from the source with the minimal source metadata value (see \aggreg{MaxSourceMetadata}).
  \item[Latest.] Returns the most recent (non-\dbnull) value. Recency may be available from another attribute, value/entity metadata or source metadata (the last case is a special case of \aggreg{MaxSourceMetadata}).
  \item[ChooseSource.] Returns a value originating from the source given as a parameter.
  \item[Vote.] Returns the most-frequently occurring  (non-\dbnull) value. Different strategies may be employed in case of tie, e.g., choosing the first or a random value.
  \item[WeightedVote.] Same as \aggreg{Vote} but each occurrence of a value is weighted by the quality of its source.
  \item[Longest, Shortest.] Returns the longest/shortest (non-\dbnull) value.
  \item[Max, Min.] Returns the maximal/minimal  (non-\dbnull) value according to an ordering of input values.
  \item[Filter.] Returns values within a given range. The minimum and/or maximum are given as parameters.
  \item[MostGeneral.] Returns the most general value according to a taxonomy or ontology.
  \item[MostSpecific.] Returns the most specific value, according to a taxonomy or ontology (if the values are on a common path in the taxonomy).
  \item[Concat.] Returns a concatenation of all values. The separator of values may be given as a parameter. Annotations such as source identifiers may be added to the result.
  \item[Constant.] Returns a constant value. The constant may be given as a parameter or be fixed (e.g. \dbnull).
  \item[CommonBeginning.] Returns the common substring at the beggining of conflicting  values.
  \item[CommonEnding.] Returns the common substring at the end of conflicting values.
  \item[TokenUnion.] Tokenizes the conflicting values and returns the union of the tokens.
  \item[TokenIntersection.] Tokenizes the conflicting values and returns the intersection of the tokens.
  \item[Avg.] Returns the average of all  (non-\dbnull) input values.
  \item[Median.] Returns the median of all  (non-\dbnull) input values.
  \item[Sum.] Returns the sum of all  (non-\dbnull) input values.
  \item[Count.] Returns the number of distinct  (non-\dbnull) 
    values.
  \item[Variance, StdDev.] Returns the variance or standard deviation of values, respectively.
  \item[ChooseCorresponding.] Returns the value that belongs to an entity (resource) whose value has already been chosen for an attribute $A$, where $A$ is given as a parameter.
  \item[ChooseDepending.] Returns the value that belongs to an entity (resource) which has a value $v$ of an attribute $A$, where $v$ and $A$ are given as parameters.
  \item[MostComplete.] Returns the  (non-\dbnull) value from the source having fewest \dbnull{}s for the respective attribute across all entities.
  \item[MostDistinguishing.] Returns the most distinguishing value among all present values for the respective attribute.
  \item[Lookup.] Returns a value by doing a lookup into the source given as a parameter, using the input values.
  \item[MostActive.] Returns the most often accessed or used value.
  \item[GlobalVote.] Returns the most-frequently occurring (non-\dbnull) value for the respective attribute among all entities in the data source.
\end{description}


\end{document}